\documentclass[aps,preprint,titlepage]{revtex4}%
\usepackage{amsfonts}
\usepackage{amsmath}
\usepackage{amssymb}
\usepackage{graphicx}%
\providecommand{\U}[1]{\protect\rule{.1in}{.1in}}

\begin{document}
\title{Modified newtonian dynamics and\textbf{ }non-relativistic ChSAS gravity}
\author{G. Rubio$\thanks{gurubio@udec.cl}$ and P. Salgado$\thanks{pasalgad@udec.cl}$}
\affiliation{Departamento de F\'{\i}sica, Universidad de Concepci\'{o}n, Casilla 160-C,
Concepci\'{o}n, Chile}

\begin{abstract}
In the context of the non-relativistic theories, a generalization of the
Chern--Weil-theorem allows us to show that\textbf{ }extended Chern--Simons
actions\textbf{ }for gravity in $d=4$ invariant under\textbf{ }some specific
non-relativistic groups lead to modified Poisson equations. In some particular
cases, these modified equations have the form of the so-called MOND approach
to gravity. The modifications could be understood as due to the effects of
dark matter. This result could leads us to think that dark matter can be
interpreted as a non-relativistic limit of dark energy.

\end{abstract}
\maketitle

\section{\textbf{Introduction}}

In Ref. \cite{grubio} it was shown that:\ $(i)$ it is possible to obtain
non-relativistic versions of both generalized Poincar\'{e} algebras
$\mathfrak{B}_{n}$ \cite{sgr,salg1,gr-bi}\ \ and generalized AdS-Lorentz
algebras \cite{soro1,soro2,soro3,salg1}. $(ii)$ Using an analogous procedure
to that used in Ref. \cite{newton}, it is possible to find\ the
non-relativistic limit of the five-dimensional Einstein--Chern--Simons gravity
which leads to a modified version of the Poisson equation.

On the other hand, in the context of the so-called extended gauge theory,
Antoniadis, Konitopoulos and Savvidy introduced background-free gauge
invariants including gauge potentials described by higher degree differential
forms \cite{sav1,sav2,sav3,sav4}. \ This construction has allowed to study in
Refs. \cite{munoz,sea,seba2017} a generalization of the Chern--Weil theorem,
which made possible to construct generalized $2n$ and $\left(  2n+2\right)
$-dimensional trangression forms, to reproduce the $(2n+2)$-dimensional
Chern--Simons forms obtained in Refs. \cite{sav3,sav4} and to show that the
$2n$-dimensional Chamseddine's topological gravity \cite{cham1,cham2,cham3}
corresponds to a Chern--Simons--Antoniadis--Savvidy (ChSAS) form. These
mathematical results were then used to study the construction of an off-shell
invariant ChSAS action for gravity in $d=4$ which is gauge quasi-invariant
under the generalized gauge transformations for the Maxwell algebra. In Ref.
\cite{ss2018} it was shown that the extended invariants found by Antoniadis,
Konitopoulos and Savvidy can also be obtained by gauging free differential algebras.

It is the purpose of this paper to consider the non-relativistic versions of
the generalized Poincar\'{e} algebra $\mathfrak{B}_{4}$ (Maxwell algebra
\cite{bacry,schr}) denoted by $\mathcal{G}\mathfrak{B}_{4}$ and the
AdS-Lorentz algebra $AdS\mathfrak{L}_{4}$ denoted by $\mathcal{G}%
\mathfrak{L}_{4}$ to find\ the non-relativistic limit of the four dimensional
ChSAS action for gravity.

This paper is organized as follows. In Section 2 we present a short review:
$\left(  i\right)  $ on the non-relativistic versions of the Maxwell algebra
$\mathfrak{B}_{4}$ and the AdS-Lorentz,  known as galilean algebra type I
$\mathcal{G}\mathfrak{B}_{4}$ and  galilean algebra type II $\mathcal{G}%
\mathfrak{L}_{4}$ respectively; $\left(  ii\right)  $ on ChSAS gravity. In
Sections 3 and 4,\ using an analogous procedure to that used in Ref.
\cite{newton}, generalizations of the Newtonian gravity are found by gauging
the $\mathcal{G}\mathfrak{B}_{4}$ and $\mathcal{G}\mathfrak{L}_{4}$ algebras.
\ In section 5 we study the possible relations between the Newtonian gravities
found in the previous sections and the so called Modified Newtonian Dynamics
(MOND) models.\ Finally our conclusions are presented in Section 6.

\section{Non-relativistic $\mathcal{G}\mathfrak{B}_{_{4}}$ and $\mathcal{G}%
\mathfrak{L}_{_{4}}$ algebras and ChSAS gravity}

In \cite{grubio} were found the non-relativistic versions of the generalized
Poincar\'{e} algebras $\mathfrak{B}_{n}$ \cite{sgr,gr-bi} and the generalized
AdS-Lorentz algebras $AdS\mathfrak{L}_{n}$ \cite{ssep,soro1,soro2,soro3,salg1}
using a generalized In\"{o}n\"{u}--Wigner contraction.\ These non-relativistic
algebras were called generalized Galilean algebras type I and type II and
denoted by $\mathcal{G}\mathfrak{B}_{_{n}}$ and $\mathcal{G}\mathfrak{L}%
_{_{n}}$ respectively.\newline

\subsection{Galilean algebra type I $\mathcal{G}\mathfrak{B}_{4}$}

In Ref. \cite{grubio} was found in that,\ separating the spatial and temporal
components in the generators $\left\{  P_{a},J_{ab},Z_{ab}\right\}  $ of
Maxwell algebra $\mathfrak{B}_{4}$, performing the rescaling $K_{i}\rightarrow
c^{-1}J_{i0}$, $P_{i}\rightarrow R^{-1}P_{i}$, $H\rightarrow cR^{-1}%
P_{0}-c^{2}M$, $\ Z_{i}\rightarrow c^{-1}Z_{i0}$ and taking the limits
$c,R\rightarrow\infty$ (with $c$ the speed of light and $R$ is the cosmic
radius), the generators of the $\mathcal{G}\mathfrak{B}_{_{4}}$ algebra
satisfy the following non-vanishing commutation relations%
\begin{align}
\lbrack J_{ij},J_{kl}] &  =\delta_{jk}J_{il}+\delta_{il}J_{jk}-\delta
_{jl}J_{ik}-\delta_{ik}J_{jl},\nonumber\\
\lbrack J_{ij},K_{k}] &  =\delta_{kj}K_{i}-\delta_{ki}K_{j},\nonumber\\
\lbrack K_{i},P_{j}] &  =-\delta_{ij}M,\nonumber\\
\lbrack J_{ij},P_{k}] &  =\delta_{kj}P_{i}-\delta_{ki}P_{j},\nonumber\\
\lbrack K_{i},H] &  =-P_{i},\nonumber\\
\lbrack J_{ij},Z_{kl}] &  =\delta_{jk}Z_{il}+\delta_{il}Z_{jk}-\delta
_{jl}Z_{ik}-\delta_{ik}Z_{jl},\nonumber\\
\lbrack J_{ij},Z_{k}] &  =\delta_{kj}Z_{i}-\delta_{ki}Z_{j},\nonumber\\
\lbrack P_{i},H] &  =Z_{i},\nonumber\\
\lbrack Z_{ij},K_{k}] &  =\delta_{kj}Z_{i}-\delta_{ki}Z_{j},\nonumber\\
\text{Others} &  =0,
\end{align}
where $i,j,k,l=1,2,3$. This algebra correspond to a S-expansion of the
Newton--Hooke algebra with central extension \cite{sexp,sexp2,nh}, whose
representation can be obtained from the $\mathrm{SO}(3,2)$ algebra using the
gamma matrices%
\begin{equation}
\Gamma_{\mu\nu}=\frac{1}{4}\left[  \gamma_{\mu},\gamma_{\nu}\right]  ,
\end{equation}
where $\gamma_{\mu}$ satisfied the Clifford algebra\textbf{\ }$\gamma_{\mu
}\gamma_{\nu}+\gamma_{\nu}\gamma_{\mu}=2\eta_{\mu\nu}$ with $\mu
,\nu=0,1,\ldots,4$\textbf{\ } and $\eta_{\mu\nu}=\mathrm{diag}\left(
-c^{2},1,1,1,-R^{2}\right)  $. The identification $J_{ij}=\Gamma_{ij},$
$\Gamma_{i0}=cK_{i},$ $\Gamma_{i4}=RP_{i},$ $\Gamma_{04}=RP_{0}=Rc^{-1}\left(
H+c^{2}M\right)  $ and $\Gamma_{\ast}=2M$ with $\Gamma_{\ast}=\gamma_{0}%
\gamma_{1}\gamma_{2}\gamma_{3}\gamma_{4}$ gives us the commutation relations
of Newton--Hooke algebra with central extension and allows to know the
invariant tensors of generalized Galilean algebras by means of the S-expansion
procedure \cite{sexp,sexp2}.

\subsection{\textbf{Galilean algebra type }II $\mathcal{G}\mathfrak{L}_{_{4}}%
$}

The $\mathcal{G}\mathfrak{L}_{4}$ algebra corresponds to the nonrelativistic
limit of the AdS$\mathcal{L}_{4}\equiv so(D-1,1)\oplus so(D-1,2)$ algebra
\cite{grubio}. \ This algebra was introduced in Refs. \cite{soro1,soro2,soro3}%
,  reobtained from the Maxwell algebra in Ref. \cite{gomis} using a method
known as deformation of Lie algebras and later from de $AdS$ algebra in Ref.
\cite{salg1} using the so called S-expansion procedure.

In Ref. \cite{grubio} it was found that separating the spatial and temporal
components in the generators $\left\{  P_{a},J_{ab},Z_{ab}\right\}  $ of
AdS$\mathcal{L}_{4}$ algebra, performing the rescaling $K_{i}\longrightarrow
c^{-1}J_{i0}$, $P_{i}\longrightarrow R^{-1}P_{i}$, $H\longrightarrow
cR^{-1}P_{0}-c^{2}M\,$, $Z_{i}\longrightarrow c^{-1}Z_{i0}$ and taking the
limit $c,R\rightarrow\infty$, \ the generators of the $\mathcal{G}%
\mathfrak{L}_{_{4}}$ algebra satisfy the following non-vanishing commutation relations%

\begin{align*}
\lbrack J_{ij},J_{kl}]  &  =\delta_{jk}J_{il}+\delta_{il}J_{jk}-\delta
_{jl}J_{ik}-\delta_{ik}J_{jl},\\
\lbrack J_{ij},K_{k}]  &  =\delta_{kj}K_{i}-\delta_{ki}K_{j},\\
\lbrack K_{i},P_{j}]  &  =-\delta_{ij}M,\\
\lbrack J_{ij},P_{k}]  &  =\delta_{kj}P_{i}-\delta_{ki}P_{j},\\
\lbrack K_{i},H]  &  =-P_{i},\\
\lbrack J_{ij},Z_{kl}]  &  =\delta_{jk}Z_{il}+\delta_{il}Z_{jk}-\delta
_{jl}Z_{ik}-\delta_{ik}Z_{jl},\\
\lbrack J_{ij},Z_{k}]  &  =\delta_{kj}Z_{i}-\delta_{ki}Z_{j},
\end{align*}

\begin{align}
\lbrack P_{i},H] &  =Z_{i},\nonumber\\
\lbrack Z_{ij},K_{k}] &  =\delta_{kj}Z_{i}-\delta_{ki}Z_{j},\nonumber\\
\lbrack Z_{ij},Z_{kl}] &  =\delta_{jk}Z_{il}+\delta_{il}Z_{jk}-\delta
_{jl}Z_{ik}-\delta_{ik}Z_{jl},\nonumber\\
\lbrack Z_{ij},Z_{k}] &  =\delta_{kj}Z_{i}-\delta_{ki}Z_{j},\nonumber\\
\lbrack Z_{ij},P_{k}] &  =\delta_{kj}P_{i}-\delta_{ki}P_{j},\nonumber\\
\lbrack Z_{i},P_{j}] &  =-\delta_{ij}M,\nonumber\\
\lbrack Z_{i},H] &  =-P_{i}.
\end{align}
This algebra can be also written as the direct sum $\mathcal{G}\mathfrak{L}%
_{4}=NH\oplus E(3),$ where $NH$ is the Newton-Hooke with central extension and
$E(3)$ is the Euclidean algebra in three dimensions. \ In fact, carrying out
the base change%

\[
\tilde{K}_{i}=K_{i}-Z_{i}\text{, \ \ }\tilde{Z}_{ij}=J_{ij}-Z_{ij},
\]
in the $\mathcal{G}\mathfrak{L}_{4}$ algebra $(3)$, we find that the only non
vanishing commutator are:

\begin{description}
\item[(a)] the Newton-Hooke algebra with central extension%
\begin{align}
\lbrack J_{ij},J_{kl}] &  =\delta_{jk}J_{il}+\delta_{il}J_{jk}-\delta
_{jl}J_{ik}-\delta_{ik}J_{jl},\nonumber\\
\lbrack J_{ij},P_{k}] &  =\delta_{kj}P_{i}-\delta_{ki}P_{j},\nonumber\\
\lbrack J_{ij},Z_{k}] &  =\delta_{kj}Z_{i}-\delta_{ki}Z_{j},\nonumber\\
\lbrack Z_{i},P_{j}] &  =-\delta_{ij}M,\nonumber\\
\lbrack Z_{i},H] &  =-P_{i},\nonumber\\
\lbrack P_{i},H] &  =Z_{i}.\label{nh}%
\end{align}

\item[(b)] the Euclidian algebra in three dimensions $E(3)$%
\begin{align}
\lbrack\tilde{Z}_{ij},\tilde{Z}_{kl}] &  =\delta_{jk}\tilde{Z}_{il}%
+\delta_{il}\tilde{Z}_{jk}-\delta_{jl}\tilde{Z}_{ik}-\delta_{ik}\tilde{Z}%
_{jl},\nonumber\\
\lbrack\tilde{Z}_{ij},\tilde{K}_{k}] &  =\delta_{kj}\tilde{K}_{i}-\delta
_{ki}\tilde{K}_{j},\nonumber\\
\left[  \tilde{K}_{i},\tilde{K}_{j}\right]   &  =0.\label{e3}%
\end{align}

\item[(c)] $J_{ij}$ conmutators are given by%
\begin{align}
\lbrack J_{ij},P_{k}] &  =\delta_{kj}P_{i}-\delta_{ki}P_{j},\nonumber\\
\lbrack J_{ij},Z_{k}] &  =\delta_{kj}Z_{i}-\delta_{ki}Z_{j},\nonumber\\
\lbrack J_{ij},\tilde{Z}_{kl}] &  =\delta_{jk}\tilde{Z}_{il}+\delta_{il}%
\tilde{Z}_{jk}-\delta_{jl}\tilde{Z}_{ik}-\delta_{ik}\tilde{Z}_{jl},\nonumber\\
\lbrack J_{ij},\tilde{K}_{k}] &  =\delta_{kj}\tilde{K}_{i}-\delta_{ki}%
\tilde{K}_{j},\nonumber\\
\text{Others} &  =0.\label{e4}%
\end{align}
From where we can see that the Newton Hooke algebra with central extension is
subalgebra of $\mathcal{G}\mathfrak{L}_{4}$ which correspond to the
non-relativistic limit of $AdS$ algebra$.$ It is interesting to note that in
(\ref{e3}) the generator $\tilde{Z}_{ij}$ corresponds to a rotation in $so(3)$
and $\tilde{K}_{i}$ is a translation in $R%
{{}^3}%
$. On the other hand, in (\ref{e4}) $J_{ij}$ corresponds to a rotation in
$so(3)$, $\tilde{Z}_{i}$ is a boost, $M$ corresponds to the center of the
algebra and $P_{i}$, $H$ are space and time translation operators respectively.
\end{description}

\subsection{ Pontryagin--Chern and ChSAS forms}

The idea of extending the Yang--Mills fields to higher rank tensor gauge
fields was used in Refs. \cite{sav2,sav3,sav4} to construct gauge invariant
and metric independent forms in higher dimensions. These forms are analogous
to the Pontryagin--Chern forms in Yang--Mills gauge theory and are given by,%

\begin{align}
\Gamma_{2n+3} &  =\langle F^{n},F_{3}\rangle=\mathrm{d}\mathfrak{C}%
_{\mathrm{ChSAS}}^{\left(  2n+2\right)  },\nonumber\\
\Gamma_{2n+4} &  =\langle F^{n},F_{4}\rangle=\mathrm{d}\mathfrak{C}%
_{\mathrm{ChSAS}}^{\left(  2n+3\right)  },\nonumber\\
\Xi_{2n+6} &  =\langle F^{n},F_{6}\rangle+n\langle F^{n-1},F_{4}^{2}%
\rangle=\mathrm{d}\mathfrak{C}_{\mathrm{ChSAS}}^{\left(  2n+5\right)
},\nonumber\\
\Upsilon_{2n+8} &  =\langle F^{n},F_{8}\rangle+3n\langle F^{n-1},F_{4}%
,F_{6}\rangle+n(n-1)\langle F^{n-2},F_{4}^{\text{ }3}\rangle=\mathrm{d}%
\mathfrak{C}_{\mathrm{ChSAS}}^{\left(  2n+7\right)  },
\end{align}
where $F_{3},F_{4},F_{6},F_{8}$ are the field-strength tensors for the gauge
fields $A_{2},$ $A_{3}$, $A_{5},$ and $A_{7}$ respectively  and where
$\mathfrak{C}_{\mathrm{ChSAS}}^{\left(  2n+2\right)  },$ $\mathfrak{C}%
_{\mathrm{ChSAS}}^{\left(  2n+3\right)  }$, $\mathfrak{C}_{\mathrm{ChSAS}%
}^{\left(  2n+5\right)  },$ $\mathfrak{C}_{\mathrm{ChSAS}}^{\left(
2n+7\right)  }$  are the corresponding Chern--Simons-Antoniads Savvidy forms
which are given by%
\begin{align}
\mathfrak{C}_{\mathrm{ChSAS}}^{\left(  2n+2\right)  } &  =\left\langle
F^{n},A_{2}\right\rangle +\mathrm{d}\varphi_{2n+1},\nonumber\\
\mathfrak{C}_{\mathrm{ChSAS}}^{\left(  2n+3\right)  } &  =\left\langle
F^{n},A_{3}\right\rangle +\mathrm{d}\varphi_{2n+2},\nonumber\\
\mathfrak{C}_{\mathrm{ChSAS}}^{\left(  2n+5\right)  } &  =\left\langle
F^{n},A_{5}\right\rangle +n\left\langle F^{n-1},F_{4},A_{3}\right\rangle
,\nonumber\\
\mathfrak{C}_{\mathrm{ChSAS}}^{\left(  2n+7\right)  } &  =\left\langle
F^{n},A_{7}\right\rangle +n(n-1)\left\langle F_{4},F_{4},A_{3},F^{n-2}%
\right\rangle +n\left\langle F_{6},A_{3},F^{n-1}\right\rangle +2n\left\langle
F_{4},A_{5},F^{n-1}\right\rangle .
\end{align}

In Ref. \cite{ss2018} were shown\ that the ChSAS invariants found in Refs.
\cite{sav2,sav3,sav4} can be constructed from a algebraic structure known as
free differential algebra (FDA).\ 

\subsection{Four-dimensional ChSAS gravity}

In Ref. \cite{munoz} was constructed generalized $\left(  2n+2\right)
$-dimensional trangression forms and reproduced the so called $(2n+2)$%
-dimensional ChSAS forms obtained in Refs. \cite{sav3,sav4}. These
mathematical results were used to construct a four-dimensional action for
gravity in $d=4$%
\begin{equation}
S_{\mathrm{ChAS}}=\int_{M^{4}}\mathcal{L}_{\mathrm{ChSAS}},
\end{equation}
whose Lagrangian $\mathcal{L}_{\mathrm{ChSAS}}=\left\langle F,A_{2}%
\right\rangle \equiv\left\langle F,B\right\rangle $ is a ChSAS form.

\section{Newtonian ChSAS gravity for the $\mathcal{G}\mathfrak{B}_{4}$
algebra}

In Ref. \cite{newton} was shown how the Newton--Cartan formulation of
Newtonian gravity can be obtained from gauging the Bargmann algebra. In Refs.
\cite{munoz} it was shown that the gauging of $\mathfrak{B}_{4}$ allows us to
construct a four-dimensional ChSAS gravity which leads general relativity in a
certain limit. On the other hand, we have seen that the non-relativistic
version of the $\mathfrak{B}_{4}$ algebra is given by the $\mathcal{G}%
\mathfrak{B}_{_{4}}$ algebra. In this Section we show that,\ using an
analogous procedure to that used in Ref. \cite{newton}, it is possible to find
a generalization of the Newtonian gravity.

\subsection{\textbf{Gauging the }$\mathcal{G}\mathfrak{B}_{_{4}}$\textbf{
algebra}}

The one-form gauge connection $A$ valued in the $\mathcal{G}\mathfrak{B}%
_{_{4}}$ algebra is given by%
\begin{equation}
A=\frac{v}{l}\tau H+\frac{1}{l}e^{i}P_{i}+\frac{1}{vl}mM+\frac{1}{v}\omega
^{i}K_{i}+\frac{1}{2}\omega^{ij}J_{ij}+\frac{1}{2}k^{ij}Z_{ij}+\frac{1}%
{v}k^{i}Z_{i},
\end{equation}
where $\nu,l$ are parameters of dimensions velocity and length respectively.
The corresponding $2$-form curvature $F=\mathrm{d}A+\frac{1}{2}[A,A]$ is given
by%
\begin{align}
F &  =\frac{v}{l}\mathrm{d}\tau H+\frac{1}{l}(T^{i}-\omega^{i}\tau)P_{i}%
+\frac{1}{vl}(\mathrm{d}m-\omega^{i}e_{i})M+\frac{1}{v}\mathrm{D}\omega
^{i}K_{i}+\frac{1}{2}R^{ij}J_{ij}\nonumber\\
&  +\frac{1}{v}(\mathrm{D}k^{i}+\frac{v^{2}}{l^{2}}e^{i}\tau+k_{\text{ \ }%
j}^{i}\omega^{j})Z_{i}+\frac{1}{2}\mathrm{D}k^{ij}Z_{ij},
\end{align}
where $T^{i}=\mathrm{D}e^{i}$ and $R^{ij}=\mathrm{d}\omega^{ij}+\omega
^{ik}\omega_{k}^{\text{ \ }j}$. Here $\mathrm{D}$ is the covariant derivative
with respect to the $\mathrm{SO}(3)$ transformations. \ For the two-form gauge
potential $B$ we can write%
\begin{equation}
B=B^{i}P_{i}+B^{0}H+B^{(m)}M+\frac{1}{2}B^{ij}J_{ij}+G^{i}K_{i}+\frac{1}%
{2}\beta^{ij}Z_{ij}+\beta^{i}Z_{i},
\end{equation}
whose associated $3$-form curvature is given by%
\begin{align}
F_{3} &  =\mathrm{D}B=\mathrm{d}B+[A,B]\nonumber\\
&  =H^{0}H+H^{i}P_{i}+H^{(m)}M+\frac{1}{2}H^{ij}J_{ij}+L^{i}K_{i}+\frac{1}%
{2}\Theta^{ij}Z_{ij}+\Theta^{i}Z_{i},
\end{align}
where%
\begin{align}
H^{0} &  =\mathrm{d}B^{0},\nonumber\\
H^{i} &  =\mathrm{D}B^{i}+\frac{v}{l}\tau G^{i}-\frac{1}{l}B^{ij}e_{j}%
-\frac{1}{v}\omega^{i}B^{0},\nonumber\\
H^{(m)} &  =\mathrm{d}B^{(m)}+\frac{1}{l}e^{i}G_{i}-\frac{1}{v}\omega_{i}%
B^{i},\nonumber\\
H^{ij} &  =\mathrm{D}B^{ij},\nonumber\\
L^{i} &  =\mathrm{D}G^{i}-\frac{1}{v}B^{ij}\omega_{j},\nonumber\\
\Theta^{ij} &  =\mathrm{D}\beta^{ij}-k_{\text{ \ }k}^{i}B^{kj}+k_{\text{ \ }%
k}^{j}B^{ki},\nonumber\\
\Theta^{i} &  =\mathrm{D}\beta^{i}-\frac{v}{l}\tau B^{i}+\frac{1}{l}e^{i}%
B^{0}-\frac{1}{v}B^{ij}k_{j}+k^{ij}G_{j}-\frac{1}{v}\beta^{ij}\omega_{j}.
\end{align}
These equations are analogous to Eq.~(2.13) of Ref.~\cite{deauria} and
Eq.~(III.6.47) of Ref.~\cite{castell}  and therefore they are not a FDA.
However, when the condition $H^{0}=H^{i}=H^{(m)}=H^{ij}=L^{i}=\Theta
^{ij}=\Theta^{i}=0$ is imposed  we get the FDA for the fields $B^{i}%
,B^{0},B^{(m)},B^{ij},G^{i},\beta^{ij},\beta^{i}$. \ The problem now is to
express the form $B$\  in terms of the one-forms $\tau,e^{i},m,\omega
^{ij},\omega^{i},k^{ij},k^{i}$ of the non-relativistic Maxwell algebra.

To express the $2$-forms as the wedge product of the $1$-forms, we follow a
procedure developed in Refs.~\cite{ascar,deauria}. We impose\textbf{ }the
ansatz%
\begin{align}
B^{i} &  =\frac{a_{1}}{2l}\omega^{ij}e_{j}+\frac{a_{2}}{2l}\omega^{i}%
\tau+\frac{a_{3}}{2l}k^{ij}e_{j}+\frac{a_{4}}{2l}k^{i}\tau+\frac{a_{5}}%
{2v^{2}l}\omega^{i}m+\frac{a_{6}}{2v^{2}l}k^{i}m,\nonumber\\
B^{0} &  =\frac{b_{1}}{2vl}\omega^{i}e_{i}+\frac{b_{2}}{2vl}k^{i}%
e_{i},\nonumber\\
B^{(m)} &  =\frac{c_{1}}{2vl}\omega^{i}e_{i}+\frac{c_{2}}{2vl}k^{i}%
e_{i},\nonumber\\
B^{ij} &  =\frac{d_{1}}{2l^{2}}e^{i}e^{j}+\frac{d_{2}}{2}\omega_{\text{ \ }%
k}^{i}\omega^{kj}+\frac{d_{3}}{2v^{2}}\omega^{i}\omega^{j}+\frac{d_{4}}%
{2}k_{\text{ \ }k}^{i}k^{kj}+\frac{d_{5}}{2v^{2}}k^{i}k^{j}\nonumber\\
&  +\frac{d_{6}}{2}\omega_{\text{ \ }k}^{i}k^{kj}+\frac{d_{7}}{2}k_{\text{
\ }k}^{i}\omega^{kj}+\frac{d_{8}}{2v^{2}}\omega^{i}k^{j}+\frac{d_{9}}{2v^{2}%
}k^{i}\omega^{j},\nonumber\\
G^{i} &  =\frac{f_{1}v}{2l^{2}}e^{i}\tau+\frac{f_{2}}{2vl^{2}}e^{i}%
m+\frac{f_{3}}{2v}\omega_{\text{ \ }j}^{i}\omega^{j}+\frac{f_{4}}{2v}k_{\text{
\ }j}^{i}\omega^{j}+\frac{f_{5}}{2v}\omega_{\text{ \ }j}^{i}k^{j}+\frac{f_{6}%
}{2v}k_{\text{ \ }j}^{i}k^{j},\nonumber\\
\beta^{ij} &  =\frac{g_{1}}{2l^{2}}e^{i}e^{j}+\frac{g_{2}}{2}\omega_{\text{
\ }k}^{i}\omega^{kj}+\frac{g_{3}}{2v^{2}}\omega^{i}\omega^{j}+\frac{g_{4}}%
{2}k_{\text{ \ }k}^{i}k^{kj}+\frac{g_{5}}{2v^{2}}k^{i}k^{j}\nonumber\\
&  +\frac{g_{6}}{2}\omega_{\text{ \ }k}^{i}k^{kj}+\frac{g_{7}}{2}k_{\text{
\ }k}^{i}\omega^{kj}+\frac{g_{8}}{2v^{2}}\omega^{i}k^{j}+\frac{g_{9}}{2v^{2}%
}k^{i}\omega^{j},\nonumber\\
\beta^{i} &  =\frac{h_{1}v}{2l^{2}}e^{i}\tau+\frac{h_{2}}{vl^{2}}e^{i}%
m+\frac{h_{3}}{2v}\omega_{\text{ \ }j}^{i}\omega^{j}+\frac{h_{4}}{2v}k_{\text{
\ }j}^{i}\omega^{j}+\frac{h_{5}}{2v}\omega_{\text{ \ }j}^{i}k^{j}+\frac{h_{6}%
}{2v}k_{\text{ \ }j}^{i}k^{j},\label{ansatz}%
\end{align}
where $a_{i},\ldots,a_{6}$, $b_{1}$, $b_{2}$, $c_{1}$, $c_{2}$, $d_{1}%
,\ldots,d_{9}$, $f_{1},\ldots,f_{6}$, $g_{1},\ldots,g_{9}$ and $h_{1}%
,\ldots,h_{6}$ are arbitrary constants. Introducing (\ref{ansatz}) in the
corresponding FDA for the fields $B^{i},B^{0},B^{(m)},B^{ij},G^{i},\beta
^{ij},\beta^{i},$\ we find relations between these constants. These relations
lead to the following form to(\ref{ansatz})%
\begin{align}
B^{i} &  =\frac{a_{2}}{2l}\omega^{i}\tau+\frac{a_{3}}{2l}k^{ij}e_{j}%
+\frac{a_{4}}{2l}k^{i}\tau+\frac{a_{5}}{2v^{2}l}\omega^{i}m,\nonumber\\
B^{0} &  =-\frac{a_{5}+d_{3}}{2vl}\omega^{i}e_{i},\text{ \ \ }B^{(m)}%
=\frac{c_{1}}{2vl}\omega^{i}e_{i}+\frac{a_{4}}{2vl}k^{i}e_{i},\nonumber\\
B^{ij} &  =\frac{d_{3}}{2v^{2}}\omega^{i}\omega^{j},\text{ \ \ }G^{i}%
=\frac{a_{3}-h_{6}}{2l^{2}}ve^{i}\tau+\frac{a_{3}+a_{4}}{2v}k_{\text{ \ }%
j}^{i}\omega^{j},\nonumber\\
\beta^{ij} &  =\frac{d_{3}}{2l^{2}}e^{i}e^{j}+\frac{g_{2}}{2}\omega_{\text{
\ }k}^{i}\omega^{kj}+\frac{g_{3}}{2v^{2}}\omega^{i}\omega^{j}+\frac{g_{4}}%
{2}k_{\text{ \ }k}^{i}k^{kj}+\frac{d_{3}}{2v^{2}}\omega^{i}k^{j}+\frac{d_{3}%
}{2v^{2}}k^{i}\omega^{j},\nonumber\\
\beta^{i} &  =\frac{h_{1}v}{2l^{2}}e^{i}\tau-\frac{a_{5}}{vl^{2}}e^{i}%
m+\frac{g_{2}}{2v}\omega_{\text{ \ }j}^{i}\omega^{j}+\frac{h_{4}}{2v}k_{\text{
\ }j}^{i}\omega^{j}+\frac{h_{6}}{2v}k_{\text{ \ }j}^{i}k^{j}.\label{ansatz1}%
\end{align}
There are $14$ arbitrary constants in the FDA expansion in terms of $1$-forms;
the fields given by Eqs. (\ref{ansatz1}) represent the most general solution
that can be built with the fields $\tau,e^{i},m,\omega^{ij},\omega^{i}%
,k^{ij},k^{i}$. Any choice of the constants represent a solution to the FDA$.$

\subsection{Non-relativistic ChSAS Lagrangian}

Using the theorem VII.2 of Ref. \cite{sexp} it is possible to show that the
invariant tensors for $\mathcal{G}\mathfrak{B}_{4}$ are given by%
\begin{equation}
\left\langle J_{ij}K_{k}\right\rangle =\alpha_{0}v\varepsilon_{ijk},\text{
\ }\left\langle J_{ij}Z_{k}\right\rangle =\alpha_{2}v\varepsilon_{ijk},\text{
\ }\left\langle K_{i}Z_{kl}\right\rangle =\alpha_{2}v\varepsilon_{ikl},
\end{equation}
being $\alpha_{0}$ and $\alpha_{2}$ arbitrary constants. The ChSAS Lagrangian
is given then by the following Chern-Simons form%
\begin{equation}
\mathfrak{C}_{\mathrm{ChSAS}}^{\left(  4\right)  }=\left\langle
F,B\right\rangle
\end{equation}
whose explicit form is%
\begin{align}
\mathfrak{C}_{\mathrm{ChSAS}}^{\left(  4\right)  } &  =\frac{\alpha_{0}}%
{2}\varepsilon_{ikl}\mathrm{D}\omega^{i}B^{kl}+\frac{\alpha_{2}}{2}%
\varepsilon_{ikl}\mathrm{D}\omega^{i}\beta^{kl}+\frac{\alpha_{0}}%
{2}v\varepsilon_{ijk}R^{ij}G^{k}+\frac{\alpha_{2}}{2}v\varepsilon_{ijk}%
R^{ij}\beta^{k}\nonumber\\
&  +\frac{\alpha_{2}}{2}v\varepsilon_{ijk}\mathrm{D}k^{ij}G^{k}+\frac
{\alpha_{2}}{2}\varepsilon_{ikl}\mathrm{D}k^{i}B^{kl}+\frac{\alpha_{2}}%
{2}\frac{v^{2}}{l^{2}}\varepsilon_{ikl}e^{i}\tau B^{kl}+\frac{\alpha_{2}}%
{2}\varepsilon_{ikl}k_{\text{ \ }j}^{i}\omega^{j}B^{kl}.\label{ChS1}%
\end{align}
Introducing the FDA expansion given by Eqs. (\ref{ansatz1}) in (\ref{ChS1}),
we find that when $v,l\rightarrow\infty$ we find that the non-relativistic
ChSAS Lagrangian for the $\mathcal{G}\mathfrak{B}_{4}$ algebra takes the form%
\begin{align}
\mathcal{L}_{\text{\textrm{NR-ChSAS}}}^{(4)} &  =\frac{\alpha_{2}}%
{2}\varepsilon_{ijk}R^{ij}\left(  \frac{h_{1}}{2}\frac{v^{2}}{l^{2}}e^{k}%
\tau+\frac{g_{2}}{2}\omega_{\text{ \ }l}^{k}\omega^{l}+\frac{h_{4}}%
{2}k_{\text{ \ }l}^{k}\omega^{l}+\frac{h_{6}}{2}k_{\text{ \ }l}^{k}%
k^{l}\right)  \nonumber\\
&  +\frac{\alpha_{2}}{2}\frac{g_{2}}{2}\varepsilon_{ijk}D\omega^{i}%
\omega_{\text{ \ }m}^{j}\omega^{mk}+\frac{\alpha_{2}}{2}\frac{g_{4}}%
{2}\varepsilon_{ijk}D\omega^{i}k_{\text{ \ }m}^{j}k^{mk}\nonumber\\
&  +\frac{\alpha_{0}}{2}\varepsilon_{ijk}R^{ij}\left(  \frac{a_{3}-h_{6}}%
{2}\frac{v^{2}}{l^{2}}e^{k}\tau+\frac{a_{3}+a_{4}}{2}k_{\text{ \ }l}^{k}%
\omega^{l}\right)  \nonumber\\
&  +\frac{\alpha_{2}}{2}\varepsilon_{ijk}Dk^{ij}\left(  \frac{a_{3}-h_{6}}%
{2}\frac{v^{2}}{l^{2}}e^{k}\tau+\frac{a_{3}+a_{4}}{2}k_{\text{ \ }l}^{k}%
\omega^{l}\right)  .\label{ChS2}%
\end{align}

In presence of matter, the complete Lagrangian of the theory is given by%
\begin{equation}
\mathcal{L}=\mathcal{L}_{\text{\textrm{NR-ChSAS}}}^{(4)}+\mathbb{\kappa
}\mathcal{L}_{M}.
\end{equation}
The variation of  $\mathcal{L}$ leads to the following equations of motion
\begin{equation}
\frac{\alpha_{2}h_{1}+\alpha_{0}(a_{3}-h_{6})}{4}\frac{v^{2}}{l^{2}%
}\varepsilon_{ijk}R^{ij}\tau+\frac{\alpha_{2}(a_{3}-h_{6})}{4}\frac{v^{2}%
}{l^{2}}\varepsilon_{ijk}Dk^{ij}\tau+\kappa\frac{\delta\mathcal{L}_{M}}{\delta
e^{k}}=0,
\end{equation}%
\begin{align}
&  \frac{\alpha_{2}g_{2}}{4}\varepsilon_{ijk}R^{ij}\omega_{\text{ \ }l}%
^{k}+\frac{\alpha_{2}h_{4}+\alpha_{0}(a_{3}+a_{4})}{4}\varepsilon_{ijk}%
R^{ij}k_{\text{ \ }l}^{k}-\frac{\alpha_{2}g_{2}}{4}\varepsilon_{ijk}%
\mathrm{D}\omega^{ij}\omega_{\text{ \ }l}^{k}\nonumber\\
&  +\frac{\alpha_{2}(2a_{3}+2a_{4}-g_{4})}{2}\varepsilon_{ijk}\mathrm{D}%
k^{ij}k_{\text{ \ }l}^{k}+\kappa\frac{\delta\mathcal{L}_{M}}{\delta\omega^{l}%
}=0,
\end{align}%
\begin{equation}
\frac{\alpha_{2}h_{6}}{4}\varepsilon_{ijk}R^{ij}k_{\text{ \ }l}^{k}%
+\kappa\frac{\delta\mathcal{L}_{M}}{\delta k^{l}}=0,
\end{equation}%
\begin{align}
&  \frac{\alpha_{2}}{4}\varepsilon_{ijk}\left(  \frac{h_{1}v^{2}}{l^{2}}%
T^{k}\tau+g_{2}\mathrm{D}\omega_{\text{ \ }l}^{k}\omega^{l}-g_{2}%
\omega_{\text{ \ }l}^{k}\mathrm{D}\omega^{l}+h_{4}\mathrm{D}k_{\text{ \ }%
l}^{k}\omega^{l}-h_{4}k_{\text{ \ }l}^{k}\mathrm{D}\omega^{l}\right)
\nonumber\\
&  +\frac{\alpha_{2}}{4}\varepsilon_{ijk}\left(  h_{6}\mathrm{D}k_{\text{
\ }l}^{k}k^{l}-h_{6}k_{\text{ \ }l}^{k}\mathrm{D}k^{l}\right)  -\frac
{\alpha_{2}g_{2}}{8}\varepsilon_{ijk}\omega_{l}\omega_{\text{ \ }f}^{k}%
\omega^{lf}+\frac{\alpha_{2}g_{2}}{4}\varepsilon_{ijk}\mathrm{D}\omega
^{l}\omega_{\text{ \ }l}^{k}\nonumber\\
&  +\frac{\alpha_{0}}{4}\varepsilon_{ijk}\left(  (a_{3}-h_{6})\frac{v^{2}%
}{l^{2}}T^{k}\tau+(a_{3}+a_{4})\mathrm{D}k_{\text{ \ }l}^{k}\omega^{l}%
-(a_{3}+a_{4})k_{\text{ \ }l}^{k}\mathrm{D}\omega^{l}\right)  \nonumber\\
&  +\frac{\alpha_{2}}{4}\varepsilon_{ijk}k_{l}^{\text{ \ }k}\left(  \left(
a_{3}-h_{6}\right)  \frac{v^{2}}{l^{2}}e^{l}\tau+\left(  a_{3}+a_{4}\right)
k_{\text{ \ }f}^{l}\omega^{f}\right)  -\frac{\alpha_{2}g_{4}}{8}%
\varepsilon_{ijk}\omega_{l}k_{\text{ \ }f}^{k}k^{lf}+\kappa\frac
{\delta\mathcal{L}_{M}}{\delta\omega^{ij}}=0,
\end{align}%
\begin{align}
&  \frac{\alpha_{2}}{4}\varepsilon_{ijk}R^{kl}\left(  h_{4}\omega_{l}%
+h_{6}k_{l}\right)  +\frac{\alpha_{2}g_{4}}{2}\varepsilon_{ijk}\mathrm{D}%
\omega_{l}k^{kl}+\frac{\alpha_{0}\left(  a_{3}+a_{4}\right)  }{4}%
\varepsilon_{ijk}R^{kl}\omega_{l}+\frac{\alpha_{2}\left(  a_{3}+a_{4}\right)
}{4}\varepsilon_{ijk}\mathrm{D}k^{kl}\omega_{l}\nonumber\\
&  +\frac{\alpha_{2}}{4}\varepsilon_{ijk}\left(  \left(  a_{3}-h_{6}\right)
\frac{v^{2}}{l^{2}}T^{k}\tau+\left(  a_{3}+a_{4}\right)  \mathrm{D}k_{\text{
\ }l}^{k}\omega^{l}-\left(  a_{3}+a_{4}\right)  k_{\text{ \ }l}^{k}%
\mathrm{D}\omega^{l}\right)  +\kappa\frac{\delta\mathcal{L}_{M}}{\delta
k^{ij}}=0,
\end{align}%
\begin{equation}
\frac{v^{2}}{l^{2}}\left(  \frac{\alpha_{2}h_{1}}{4}+\alpha_{0}\frac
{a_{3}-h_{6}}{4}\right)  \varepsilon_{ijk}R^{ij}e^{k}+\alpha_{2}\frac{v^{2}%
}{l^{2}}\frac{a_{3}-h_{6}}{4}\varepsilon_{ijk}Dk^{ij}e^{k}+\kappa\frac
{\delta\mathcal{L}_{M}}{\delta\tau}=0,\label{mov}%
\end{equation}
where we have used%
\begin{equation}
T_{a}=\ast\left(  \frac{\delta\mathcal{L}_{M}}{\delta e^{a}}\right)  ,\text{
\ \ \ \ \ }T_{0}=\ast\left(  \frac{\delta\mathcal{L}_{M}}{\delta\tau}\right)
.
\end{equation}
Taking into account that
\begin{align}
\ast(T_{0})\delta\tau &  =-\det(g)\delta_{\delta}^{\sigma}T_{\text{ \ }0}%
^{0}\text{\ }\delta\tau_{\text{ \ }\sigma}^{\delta}\mathrm{d}x^{4},\nonumber\\
\varepsilon_{ijk}\mathrm{D}k^{ij}e^{k}\delta\tau &  =2\det(g)\left(
\delta_{\delta}^{\sigma}\mathrm{D}_{\alpha}k_{\text{ \ \ \ }\beta}%
^{\alpha\beta}-\mathrm{D}_{\delta}k_{\text{ \ \ \ }\beta}^{\sigma\beta
}+\mathrm{D}_{\beta}k_{\text{ \ \ \ }\delta}^{\sigma\beta}\right)  \delta
\tau_{\text{ \ }\sigma}^{\delta}\mathrm{d}x^{4},\\
\varepsilon_{ijk}R^{ij}e^{k}\delta\tau &  =\det(g)\left(  \delta_{\delta
}^{\sigma}R-2R_{\text{ \ }\delta}^{\sigma}\right)  \delta\tau_{\text{
\ }\sigma}^{\delta}\mathrm{d}x^{4},\nonumber
\end{align}
we find that the field equation (\ref{mov}) can be written as%
\begin{gather}
\left[  \left(  \frac{\alpha_{2}h_{1}}{4}+\alpha_{0}\frac{a_{3}-h_{6}}%
{4}\right)  \left(  \delta_{\sigma\delta}R-2R_{\sigma\delta}\right)
+\alpha_{2}\frac{a_{3}-h_{6}}{2}\left(  \delta_{\sigma\delta}\mathrm{D}%
_{\alpha}k_{\text{ \ \ \ }\beta}^{\alpha\beta}-\mathrm{D}_{\delta}%
k_{\sigma\text{\ \ }\beta}^{\text{ \ }\beta}+\mathrm{D}_{\beta}k_{\sigma\text{
\ }\delta}^{\text{ \ }\beta}\right)  \right.  \nonumber\\
\left.  -\frac{l^{2}}{v^{2}}\kappa\delta_{\sigma\delta}T_{\text{ \ }0}%
^{0}\right]  \det(g)\delta\tau^{\delta\sigma}\mathrm{d}x^{4}=0.\label{5}%
\end{gather}
The contraction of this equation with $g^{\sigma\delta}$ leads to
\begin{equation}
\left(  \frac{\alpha_{2}h_{1}}{4}+\alpha_{0}\frac{a_{3}-h_{6}}{4}\right)
R=\alpha_{2}\frac{a_{3}-h_{6}}{2}\mathrm{D}_{\alpha}k_{\text{ \ \ \ }\beta
}^{\alpha\beta}-\frac{2l^{2}}{v^{2}}\kappa T_{00}.\label{6'}%
\end{equation}
Taking the components $00$ of (\ref{5}) and using (\ref{6'}) we find%
\begin{align}
&  -\left(  \frac{\alpha_{2}h_{1}}{2}+\alpha_{0}\frac{a_{3}-h_{6}}{2}\right)
R_{00}\nonumber\\
+\alpha_{2}\frac{a_{3}-h_{6}}{2}\left(  2\mathrm{D}_{\alpha}k_{\text{
\ \ \ }\beta}^{\alpha\beta}-\mathrm{D}_{0}k_{0\text{ \ }\beta}^{\text{
\ }\beta}+\mathrm{D}_{\beta}k_{0\text{\ \ }0}^{\text{ \ }\beta}\right)
-\frac{3l^{2}}{v^{2}}\kappa T_{00} &  =0.\label{7}%
\end{align}
Following the procedure of Ref. \cite{newton}, we find
\begin{equation}
R_{00}=\nabla^{2}\phi,\text{ \ }g_{00}=\tau_{0}\tau_{0}=1\text{, \ \ }%
T_{00}=\rho.\label{7'}%
\end{equation}
From (\ref{7'}) and (\ref{7}) we finally obtain%
\begin{equation}
\nabla^{2}\phi=6\alpha\frac{l^{2}}{v^{2}}\kappa\rho-\alpha\beta\left(
2\mathrm{D}_{\alpha}k_{\text{ \ \ \ }\beta}^{\alpha\beta}-\mathrm{D}%
_{0}k_{0\text{ \ }\beta}^{\text{ \ }\beta}+\mathrm{D}_{\beta}k_{0\text{\ \ }%
0}^{\text{ \ }\beta}\right)  ,\label{8}%
\end{equation}
where%
\begin{equation}
\alpha=-\left[  \alpha_{2}h_{1}+\alpha_{0}(a_{3}-h_{6})\right]  ^{-1}%
,\text{\ \ \ \ \ }\beta=a_{2}(a_{3}-h_{6}).\label{8'}%
\end{equation}

\section{Mond theory connection}

The modified form of Poisson equation  (\ref{8}) suggests a possible
connection with the so-called MOND approach to gravity interactions. In fact,
the first complete MOND theory was constructed by Milgrom and Bekenstein in
Ref. \cite{mond}. It involves a modification of the Poisson equation and can
be derived from the following Lagrangian%
\[
\mathcal{L}_{\text{\textrm{Mond}}}=-%
{\displaystyle\int}
d^{3}r\left\{  \rho\varphi+\frac{a_{0}^{2}}{8\pi G}\mathfrak{F}\left[
\frac{\left(  \vec{\nabla}\varphi\right)  ^{2}}{a_{0}^{2}}\right]  \right\}  ,
\]
where $\varphi$ is the gravitational potential (meaning that for a test
particle $\vec{a}=-\vec{\nabla}\varphi$), $\rho$ denotes the matter mass
density, and $\mathfrak{F(}x^{2}\mathfrak{)}$ is an arbitrary function. The
variation of $\mathcal{L}_{\text{\textrm{Mond}}}$ with respect to $\varphi$
leads to the following field equations%
\[
\vec{\nabla}\cdot\left[  \mu\left(  \frac{\left\vert \vec{\nabla}%
\varphi\right\vert }{a_{0}}\right)  \vec{\nabla}\varphi\right]  =4\pi G\rho,
\]
with $\mu(x)=\mathfrak{F}^{^{\prime}}\mathfrak{(}x^{2})$.\ A little bit of
algebra allows us to see that this equation takes the form%
\begin{equation}
\mu(x)\nabla^{2}\varphi(x)=4\pi G\rho-\vec{\nabla}\mu(x)\cdot\vec{\nabla
}\varphi(x).\label{9}%
\end{equation}

Comparing this last equation with (\ref{8}) we can see that in some particular
cases the MOND approach to gravity could coincide with such modified Poisson
equation. In fact, if we consider the case where the $k_{\text{ \ \ \ }\beta
}^{\alpha\beta}$ field does not depend on time and its non zero components are
given by $k_{\text{ \ \ \ }0}^{i0}=-k_{\text{ \ \ \ }0}^{0i}=\sigma
(x)\delta^{ij}\partial_{j}\phi(x)$, we find%
\begin{equation}
2\mathrm{D}_{\alpha}k_{\text{ \ \ \ }\beta}^{\alpha\beta}-\mathrm{D}%
_{0}k_{0\text{ \ }\beta}^{\text{ \ }\beta}+\mathrm{D}_{\beta}k_{0\text{\ \ }%
0}^{\text{ \ }\beta}=\mathrm{D}_{i}k_{\text{ \ \ \ }0}^{i0},\label{10}%
\end{equation}
where $\mathrm{D}_{\alpha}$ is the covariant derivative. Following Ref.
\cite{newton} with $\Gamma_{\text{ \ }00}^{i}=\delta^{ij}\partial_{j}\phi(x)$
we can see that%
\begin{equation}
\mathrm{D}_{i}k_{\text{ \ \ \ }0}^{i0}=\vec{\nabla}\sigma(x)\cdot\vec{\nabla
}\phi(x)+\sigma(x)\nabla^{2}\phi(x).\label{11}%
\end{equation}
Introducing (\ref{10},\ref{11}) in (\ref{8}) we find%
\begin{equation}
\left[  1+\alpha\beta\sigma(x)\right]  \nabla^{2}\phi(x)=6\alpha\frac{l^{2}%
}{v^{2}}\kappa\rho-\alpha\beta\vec{\nabla}\sigma(x)\cdot\vec{\nabla}%
\phi(x).\label{maxf'}%
\end{equation}
Comparing (\ref{maxf'}) with (\ref{9}), we can see that if $\phi
(x)=\varphi(x)$ and $\mu(x)=1+\alpha\beta\sigma(x),$  we can see that Eq.
(\ref{maxf'}) matchs with the MOND equation (\ref{9}) for the following values
of the $\alpha$ and $\beta$%
\begin{equation}
\alpha=\frac{4\pi Gv^{2}}{6l^{2}\kappa}=\frac{v^{2}}{12l^{2}},\text{ \ }%
\beta=1.
\end{equation}

\section{Newtonian ChSAS gravity for $\mathcal{G}\mathfrak{L}_{_{4}}$ algebra}

Let us now consider the one and two forms gauge fields $A$, $B$ valued in the
$\mathcal{G}\mathfrak{L}_{_{4}}$ algebra%
\begin{align}
A &  =\frac{v}{l}\tau H+\frac{1}{l}e^{i}P_{i}+\frac{1}{vl}mM+\frac{1}{v}%
\omega^{i}K_{i}+\frac{1}{2}\omega^{ij}J_{ij}+\frac{1}{2}k^{ij}Z_{ij}+\frac
{1}{v}k^{i}Z_{i},\nonumber\\
B &  =B^{i}P_{i}+B^{0}H+B^{(m)}M+\frac{1}{2}B^{ij}J_{ij}+G^{i}K_{i}+\frac
{1}{2}\beta^{ij}Z_{ij}+\beta^{i}Z_{i}.
\end{align}

Following the same procedure used in the previous section, it is found that
the corresponding non-relativistic ChSAS Lagrangian leads to the following
generalized Poisson equation%
\begin{align}
\nabla^{2}\phi &  =-\frac{6}{\alpha_{0}(a_{4}-a_{2}+a_{3})}\frac{l^{2}}{v^{2}%
}\kappa\rho+\frac{\alpha_{2}}{\alpha_{0}}\left(  2D_{\alpha}k_{\text{
\ \ \ }\beta}^{\alpha\beta}-D_{0}k_{0\text{ \ }\beta}^{\text{ \ }\beta
}+D_{\beta}k_{0\text{\ \ }0}^{\text{ \ }\beta}\right)  \nonumber\\
&  +\frac{2\alpha_{2}}{\alpha_{0}}\left(  1+\frac{a_{3}}{(a_{4}-a_{2}+a_{3}%
)}\right)  \left(  \left(  k_{\text{ \ \ \ }\alpha}^{\alpha\gamma}%
k_{\gamma\text{ \ }\beta}^{\text{ \ }\beta}-k_{\text{ \ \ \ }\beta}%
^{\alpha\gamma}k_{\gamma\text{ \ }\alpha}^{\text{ \ }\beta}\right)  \right.
\nonumber\\
&  \left.  -\left(  k_{\text{ \ \ \ }0}^{0\gamma}k_{\gamma\text{ \ }\beta
}^{\text{ \ }\beta}-k_{\text{ \ \ \ }\beta}^{0\gamma}k_{\gamma\text{ \ }%
0}^{\text{ \ }\beta}\right)  +\left(  k_{\text{ \ \ \ }0}^{\alpha\beta
}k_{\beta\text{ \ }\alpha}^{\text{ \ }0}-k_{\text{ \ \ \ }\alpha}^{\alpha
\beta}k_{\beta\text{ \ }0}^{\text{ \ }0}\right)  \right)  .\label{7''}%
\end{align}
Comparing  (\ref{7''}) with (\ref{9}) we can see that in some particular
cases, the MOND approach to gravity could coincide with such modified Poisson
equation (\ref{7''}). If we consider again the case where the $k_{\text{
\ \ \ }\beta}^{\alpha\beta}$ field does not depend on time and its non-zero
components are given by%
\begin{equation}
k_{\text{ \ \ \ }0}^{i0}=-k_{\text{ \ \ \ }0}^{0i}=\sigma(x)\delta
^{ij}\partial_{j}\phi(x),
\end{equation}
we find $k_{\text{ \ \ \ }\alpha}^{\alpha\gamma}k_{\gamma\text{ \ }\beta
}^{\text{ \ }\beta}-k_{\text{ \ \ \ }\beta}^{\alpha\gamma}k_{\gamma\text{
\ }\alpha}^{\text{ \ }\beta}=0,$ \ $k_{\text{ \ \ \ }0}^{0\gamma}%
k_{\gamma\text{ \ }\beta}^{\text{ \ }\beta}-k_{\text{ \ \ \ }\beta}^{0\gamma
}k_{\gamma\text{ \ }0}^{\text{ \ }\beta}=0,$ $k_{\text{ \ \ \ }0}%
^{\alpha\gamma}k_{\gamma\text{ \ }\alpha}^{\text{ \ }0}-k_{\text{
\ \ \ }\alpha}^{\alpha\gamma}k_{\gamma\text{ \ }0}^{\text{ \ }0}=0$. So that,
Eq. (\ref{7''}) takes the form%
\begin{equation}
\left[  1+\beta\sigma(x)\right]  \nabla^{2}\phi(x)=6\alpha\frac{l^{2}}{v^{2}%
}\kappa\rho-\beta\vec{\nabla}\sigma(x)\cdot\vec{\nabla}\phi(x)\text{
}\label{maxf''}%
\end{equation}
where%
\begin{equation}
\alpha=-\frac{1}{\alpha_{0}(a_{4}-a_{2}+a_{3})},\text{ \ \ \ }\beta
=-\frac{\alpha_{2}}{\alpha_{0}}.
\end{equation}
Comparing (\ref{maxf''}) with (\ref{9}) we can see that if $\phi
(x)=\varphi(x)$ and $\ \mu(x)=1+\beta\sigma(x),$ then Eq. (\ref{maxf''})
matchs with the Mond equation (\ref{9}) if%
\begin{align*}
\alpha &  =\frac{4\pi Gv^{2}}{6l^{2}\kappa}=\frac{v^{2}}{12l^{2}},\\
\beta &  =1.
\end{align*}

\section{\textbf{Comments }}

In the present work we have studied the non-relativistic versions of the
generalized Poincar\'{e} algebra $\mathfrak{B}_{4}$  denoted by $\mathcal{G}%
\mathfrak{B}_{4}$ and the AdS-Lorentz algebra $AdS\mathfrak{L}_{4}$ denoted by
$\mathcal{G}\mathfrak{L}_{4}$ to find\ the non-relativistic limit of the four
dimensional ChSAS action for gravity.

We have shown that the gauging of non-relativistic algebras $\mathcal{G}%
\mathfrak{B}_{_{4}}$ and $\mathcal{G}\mathfrak{L}_{_{4}}$ permits to construct
generalizations of the Newtonian gravity which leads to modified versions of
the Poisson equation. In some particular cases, it is possible to find
relations between the generalized Newtonian gravities and the so called MOND
model. These modifications to the Poisson equation, could be compatible with
dark matter and would allow us to conjecture that dark matter could be
interpreted as the non-relativistic limit of dark energy.

\textbf{Acknowledgements:} \textit{This work was supported in part }by
universidad de Concepci\'{o}n through DIUC Grant No. 217.011.056-1.0 and in
part \textit{by FONDECYT Grants }N$_{o}$\textit{ 1180681 } from the Government
of Chile. \textit{One of the authors (}GR\textit{) was supported by grants
from Comision Nacional de Investigaci\'{o}n Cient\'{\i}fica y Tecnol\'{o}gica
}CONICYT\textit{ }N$_{o}$\textit{ }21140971\textit{ and from Universidad de
Concepci\'{o}n, Chile.}


\begin{thebibliography}{99}                                                                                               %


\bibitem {grubio}N. Gonz\'{a}lez, G. Rubio, P. Salgado and S. Salgado,
\textit{Generalized Galilean algebras and Newtonian gravity,} Phys. Lett. B
\textbf{755} (2016) 433.

\bibitem {sgr}F. Izaurieta, P.Minning, E. Rodr\'{\i}guez, A. P\'{e}rez and P.
Salgado, \textit{Standard General Relativity from Chern-Simons Gravity,} Phys.
Lett. B \textbf{678} (2009) 213.

\bibitem {salg1}P. Salgado and S. Salgado, \textit{so(D-1,1)}$\oplus
$\textit{so(D-1,2)} \textit{algebras and gravity,} Phys. Lett. B \textbf{728},
(2014) 5.

\bibitem {gr-bi}P. K. Concha, D. M. Pe\~{n}afiel, E. K. Rodriguez and P.
Salgado, \textit{Even-dimensional General Relativity from Born--Infeld
gravity}, Phys. Lett. B \textbf{725} (2013) 419.

\bibitem {soro1}D. V. Soroka and V. A. Soroka, \textit{Tensor extension of the
Poincar\'{e} algebra}, Phys. Lett. B \textbf{607} (2005) 302.

\bibitem {soro2}D. V. Soroka and V. A. Soroka, \textit{Semi-simple extension
of the (super)Poincar\'{e} algebra}, Adv. High Energy Phys. \textbf{2009}
(2009) 234147.

\bibitem {soro3}D. V. Soroka and V. A. Soroka, \textit{Semi-simple
o(N)-extended super-Poincar\'{e} algebra}, arXiv:1004.3194 [hep-th].

\bibitem {newton}R. Andringa, E. Bergshoeff, S. Panda and M. de Roo,
\textit{Newtonian Gravity and the Bargmann Algebra}, Class. \& Quant. Grav.
\textbf{28} (2011) 105011.

\bibitem {sav1}G. Savvidy, \textit{Topological mass generation in
four-dimensional gauge theory}, Phys. Lett. B \textbf{694} (2010) 65.

\bibitem {sav2}S. Konitopoulos and G. Savvidy, \textit{Extension of
Chern--Simons forms and new gauge anomalies}, J. Math. Phys. \textbf{55}
(2014) 06234.

\bibitem {sav3}I. Antoniadis and G. Savvidy, \textit{New gauge anomalies and
topological invariants in various dimensions}, Eur. Phys. J. C \textbf{72}
(2012) 2140.

\bibitem {sav4}I. Antoniadis and G. Savvidy, \textit{Extension of
Chern--Simons forms and new gauge anomalies}, Int. J. Mod. Phys. A \textbf{29}
(2014) 1450027.

\bibitem {munoz}F. Izaurieta, I. Mu\~{n}oz, P. Salgado, \textit{A
Chern--Simons gravity action in d=4}, Phys. Lett. B \textbf{750} (2015) 39.

\bibitem {sea}P. Catal\'{a}n, F. Izaurieta, P. Salgado and S. Salgado,
\textit{Topological gravity and Chern--Simons forms in d=4}, Phys. Lett. B
\textbf{751} (2015) 205.

\bibitem {seba2017}F. Izaurieta, P. Salgado, S. Salgado,
\textit{Chern-Simons-Antoniadis-Savvidy forms and standard supergravity},
Phys. Lett. B \textbf{767} (2017) 360.

\bibitem {cham1}A. H. Chamseddine, \textit{Topological gravity and
supergravity in various dimensions}, Nucl. Phys. B \textbf{346} (1990) 213.

\bibitem {cham2}A. H. Chamseddine, \textit{Topological Gauge Theory of Gravity
in Five-dimensions and All Odd Dimensions}, Phys. Lett. B \textbf{233} (1989) 291.

\bibitem {cham3}A. H. Chamseddine, Nucl. Phys. B \textbf{340} (1990) 595.

\bibitem {ss2018}P. Salgado, S. Salgado, \textit{Extended gauge theory and
gauged free differential algebras}, Nucl. Phys. B \textbf{926} (2018) 179.

\bibitem {bacry}H. Bacry, P. Combe, J.L. Richard, \textit{Group-theoretical
analysis of elementary particles in an external electromagnetic field. 1. the
relativistic particle in a constant and uniform field,} Nuovo Cimento. A
\textbf{67} (1970) 267.

\bibitem {schr}R. Schr\"{a}der, \textit{The Maxwell Group and the Quantum
Theory of Particles in Classical Homogeneous Electromagnetic Fields},
Fortschr. Phys. \textbf{20} (1972) 701.

\bibitem {ssep}J. D\'{\i}az, O. Fierro, F. Izaurieta, N. Merino, E.
Rodr\'{\i}guez, P. Salgado and O. Valdivia, \textit{A generalized action for
(2+1)-dimensional Chern--Simons gravity}, J. Phys. A: Math. Theor.
\textbf{45}, (2012) 255207.

\bibitem {sexp}F. Izaurieta, E. Rodr\'{\i}guez and P. Salgado,
\textit{Expanding Lie (super) algebras through abelian semigroups}, J. Math.
Phys. \textbf{47} (2006) 123512.

\bibitem {sexp2}F. Izaurieta, E. Rodr\'{\i}guez and P. Salgado,\textit{ The
Extended Cartan Homotopy Formula and a Subspace Separation Method for
Chern-Simons Theory}, Lett. Math. Phys. \textbf{80} (2007) 127.

\bibitem {nh}Y. Tian, H. Y. Guo, C. G.Huang, Z. Xu and B. Zhou,
\textit{Mechanics and Newton-Cartan-like gravity on the Newton-Hooke
space-time}, Phys. Rev. D \textbf{71} (2005) 044030.

\bibitem {deauria}R. D'Auria and P. Fr\'{e}, \textit{Geometric Supergravity in
d = 11 and Its Hidden Supergroup}, Nucl. Phys. B \textbf{201} (1982) 101.

\bibitem {castell}L. Castellani, R. D'Auria and P. Fr\'{e},
\textit{Supergravity and Superstring. A Geometric Perspective,} World
Scientific 1991.

\bibitem {ascar}I. Bandos, J.A. de Azc\'{a}rraga, M. Pic\'{o}n and O. Varela,
\textit{On the formulation of D = 11 supergravity and the composite nature of
its three-form gauge field}, Ann. Phys. \textbf{317} (2005) 238.

\bibitem {salg2}A. P\'{e}rez, P. Minning and P. Salgado,
\textit{Eleven-dimensional CJS supergravity and the D'Auria-Fre group}, Phys.
Lett. B \textbf{660} (2008) 407.

\bibitem {mond}J. D. Bekenstein and M. Milgrom, \textit{Does the missing mass
problem signal the breakdown of Newtonian gravity?} Astrophys. J. \textbf{286}
(1984) 7.

\bibitem {gomis}J. Gomis, K. Kamimura, J. Lukierski, \textit{Deformations of
Maxwell algebra and their Dynamical Realizations}, JHEP \textbf{08} (2009)
039, arXiv:0906.4464 [hep-th].
\end{thebibliography}
\end{document}